\documentclass[10pt,letterpaper]{article}
\usepackage{opex3_dm}

\begin{document}

\title{Optical Vectorial Vortex Coronagraphs\\ 
using Liquid Crystal Polymers:\\
theory, manufacturing and laboratory demonstration}

\author{D. Mawet$^{1}$, E. Serabyn$^{1}$, K. Liewer$^{1}$, Ch. Hanot$^{1,3}$, S. McEldowney$^{2}$, D. Shemo$^{2}$, and N. O'Brien$^{2}$}

\address{$^1$Jet Propulsion Laboratory, California Institute of Technology, 4800 Oak Grove Dr, Pasadena, CA 91109, USA}
\address{$^2$JDSU, 2789 Northpoint Parkway, Santa Rosa, CA 95407, USA}
\address{$^3$Institut d'Astrophysique et de G\'eophysique de Li\`ege, Universit\'e de Li\`ege, All\'ee du 6 Ao\^ut, B-4000 Li\`ege, Belgium}

\email{Dimitri.Mawet@jpl.nasa.gov} 



\begin{abstract}
In this paper, after briefly reviewing the theory of vectorial vortices, we describe our technological approach to generating the necessary phase helix, and report results obtained with the first optical vectorial vortex coronagraph (OVVC) in the laboratory. To implement the geometrical phase ramp, we make use of Liquid Crystal Polymers (LCP), which we believe to be the most efficient technological path to quickly synthesize optical vectorial vortices of virtually any topological charge. With the first prototype device of topological charge 2, a maximum peak-to-peak attenuation of  $1.4\times 10^{-2}$ and a residual light level of $3\times 10^{-5}$ at an angular separation of 3.5 $\lambda/d$ (at which point our current noise floor is reached) have been obtained at a wavelength of 1.55 $\mu$m. These results demonstrate the validity of using space-variant birefringence distributions to generate a new family of coronagraphs usable in natural unpolarized light,  opening a path to high performance coronagraphs that are achromatic and have low-sensitivity to low-order wavefront aberrations.
\end{abstract}



\section{Introduction}

Coronagraphy, a means of suppressing bright starlight so as to search for faint companions or circumstellar material, has garnered much interest in the past decade. A number of new optimized coronagraph concepts has been considered for dedicated exoplanet-finding projects at both ground- and space-based observatories (see the rather exhaustive survey of \cite{Guyon2006}). Four groups of coronagraphs can be identified. In the pupil plane coronagraph family, there are: (a) amplitude apodization (via either intensity modification, or pupil shaping), and the phase induced amplitude apodization (PIAA) approach, in which the mirrors are warped to reshape the distribution of light in the pupil to yield an apodized shape, (b) pupil plane phase apodization, and in the focal plane coronagraph family, there are (c) amplitude masks (classical Lyot coronagraphs and band-limited profiles), and (d) phase mask coronagraphs of various types. Unlike ``Lyot family'' amplitude coronagraphs, focal-plane phase masks such as the four-quadrant phase-mask (FQPM, see \cite{Rouan2000}, and \cite{Rouan2007} for a recent review on the FQPM) and optical vortex coronagraphs \cite{Mawet2005a,Foo2005,Jenkins2008,Swartzlander2008} have the great advantage of providing a very small inner working angle (IWA), which is one way of making use of smaller telescope apertures.

The optical vortex coronagraph family is a promising and nearly ideal phase mask solution, proposed almost concurrently by \cite{Mawet2005a} and \cite{Foo2005}, and confirmed later in \cite{Guyon2006}. An optical vortex is a phase singularity in an optical field, a point of zero intensity, resulting from a phase screw dislocation of the form $e^{il\theta}$, with $l$ being the so-called topological charge\footnote{The topological charge of a vortex is defined according to the number of $2\pi$ radians that the phase of the scalar wave accumulates along a closed path surrounding the phase singularity
\begin{equation}\label{top}
l=\frac{1}{2\pi}\oint \nabla \phi ds
\end{equation}
where the integration path encircles the central singularity.}, and $\theta$ the azimuthal coordinate. This anomaly forces the intensity to vanish by a total destructive interference, creating a dark core. This dark core propagates and is conserved along the optical axis \cite{Rozas97,Niv06b}. Whether a dark core is created in the pupil or focal plane of a telescope will determine the way it further evolves. In \cite{Mawet2005a}, we demonstrated that the creation of a vortex of any even topological charge $l$ at the focal plane can lead to a perfect coronagraph in the ideal case (no further phase aberrations in the system). This result has been confirmed later on by \cite{Foo2005,Jenkins2008}. Even if  ``optical vortex'' rigorously refers to a property of an optical beam, for the sake of simplicity we will use this denomination in the following to refer either to the optical property or the optical element that induces it. There are two kinds of optical vortices (and thus of elements that induce them):
\begin{itemize}
\item [-] scalar optical vortex, based on the longitudinal phase delay (which applies to both polarizations identically), and implemented by a structural helical phase ramp \cite{Foo2005,Swartzlander2008};
\item [-] vectorial optical vortex, based on the ``geometrical'' or Pancharatnam phase \cite[see Appendix]{Berry87}, obtained by manipulating the transverse polarization state of the light with space-variant birefringent elements. 
\end{itemize}
Both approaches are still technically immature, but a scalar vortex optical element has already seen use on a small telescope \cite{Swartzlander2008}. Here we report laboratory results with the first prototype vectorial vortex coronagraph. After several years of effort toward annular groove vortex masks \cite{Mawet2005a}, a new promising technique has been identified to manufacture the vortex phase profile. This technology is based on birefringent liquid crystal polymers (LCP) which are oriented and hardened in a circular pattern. We show in this paper that the liquid crystal polymer optical vectorial vortex coronagraph (LCP OVVC) is a viable approach to generating the phase helix, with a promising technological path to very high performance coronagraphic masks, both in the near-infrared, where they could be used on current adaptive optics corrected ground-based telescopes to augment high contrast imaging capabilities, and in the optical, where these developments can be applied to more ambitious space-based coronagraph missions. We note that because of their small inner working angle, and lack of  ``features'' anywhere on the mask (except at the very central singularity), their discovery space is close to maximal in terms of search area for faint companions (no dead zones).

The paper is organized as follows: the second section is devoted to a condensed theoretical description of the optical vectorial vortex, as opposed to the optical ``scalar'' vortex. The third section describes the technology chosen to synthesize them as well as the manufacturing process of the first prototype. In the fourth section, the lab setup and measurement methodology will be presented. In Section 5, we will present the coronagraphic measurements results, that will subsequently be put in perspective in Section 6, before concluding in Section 7.

\section{Vectorial vortex as opposed to scalar vortex}

Making the helical phase profile $e^{il\theta}$ of optical vortices is difficult because of the accuracy required in the phase profile control. To build a scalar vortex, i.e. by acting on the longitudinal phase delay, one has to directly etch the helix in a piece of glass. The total etched  thickness $t$ must be precise to a tiny fraction $x$ of $\lambda/n$, $\lambda$ being the wavelength of the impinging light and $n$ the refractive index of the glass, respectively. In the most demanding applications such as Earth-like planet detection, one would have to structurally control the phase ramp profile at the nanometer level, which is very challenging, even for state-of-the-art etching techniques inherited from the micro-electronics industry (e-beam lithography and reactive ion etching). Moreover, the induced phase profile using this approach, $\Delta \phi = 2\pi/\lambda \times (n (\lambda)-1) \times t$, is very chromatic, with a hyperbolic dependence to the wavelength of light and subject to the glass refractive index dispersion.
\begin{figure}[!t]
  \centering
  \includegraphics[scale=0.5]{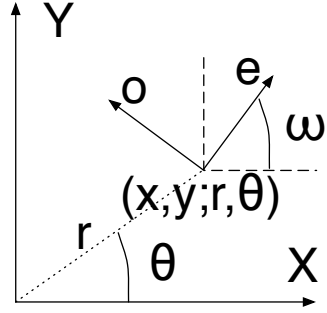}
  \caption{Space-variant birefringent medium vectorial analysis. $\textbf{X}$ and $\textbf{Y}$ are the unit vectors of the chosen cartesian basis. $o$ and $e$, are the local $ordinary$ and $extraordinary$ optical axis unit vectors of the birefringent medium. \label{spvarsch}}
\end{figure}

\begin{figure}[!t]
  \centering
\includegraphics[scale=0.4]{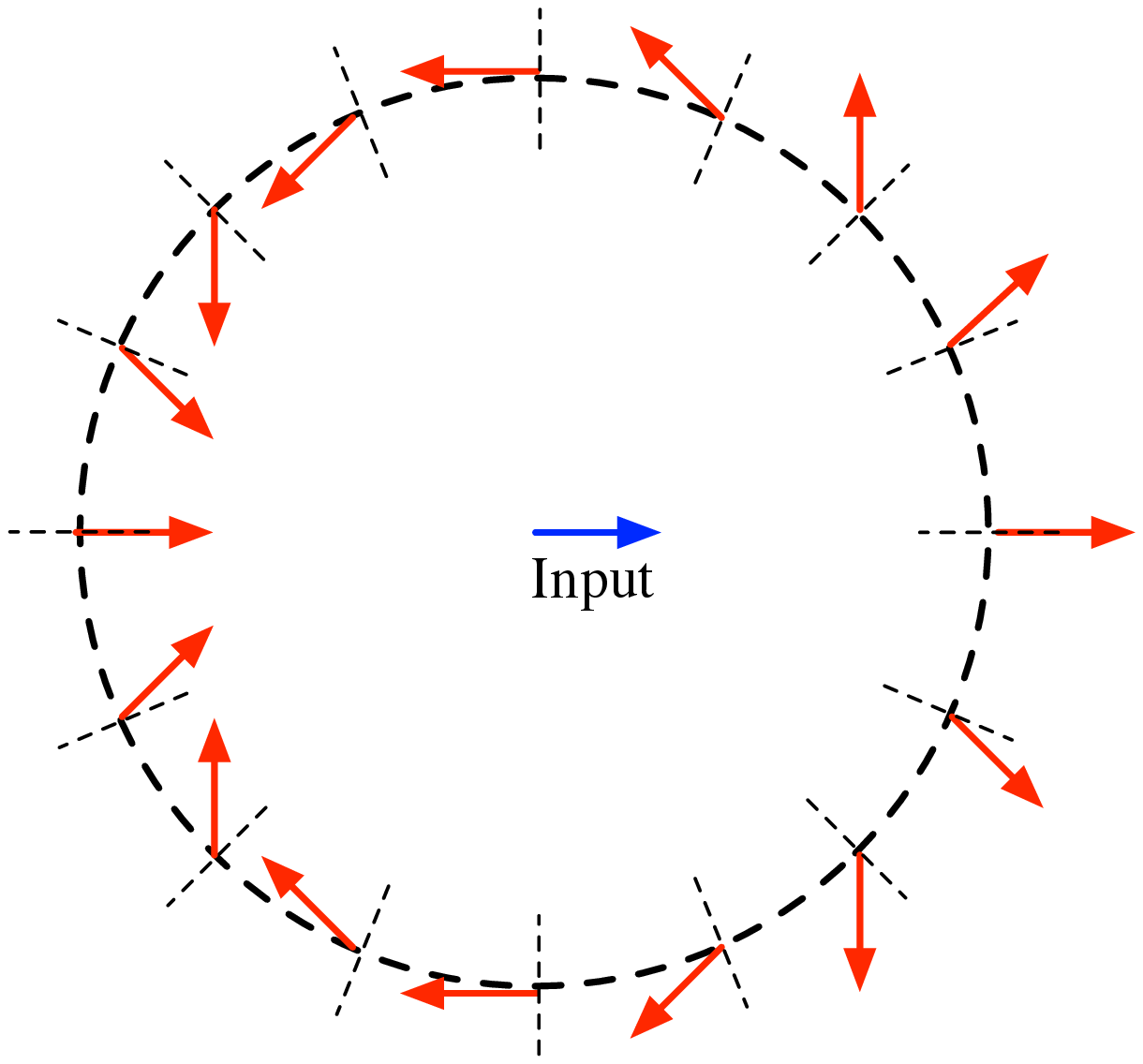}\includegraphics[scale=0.5]{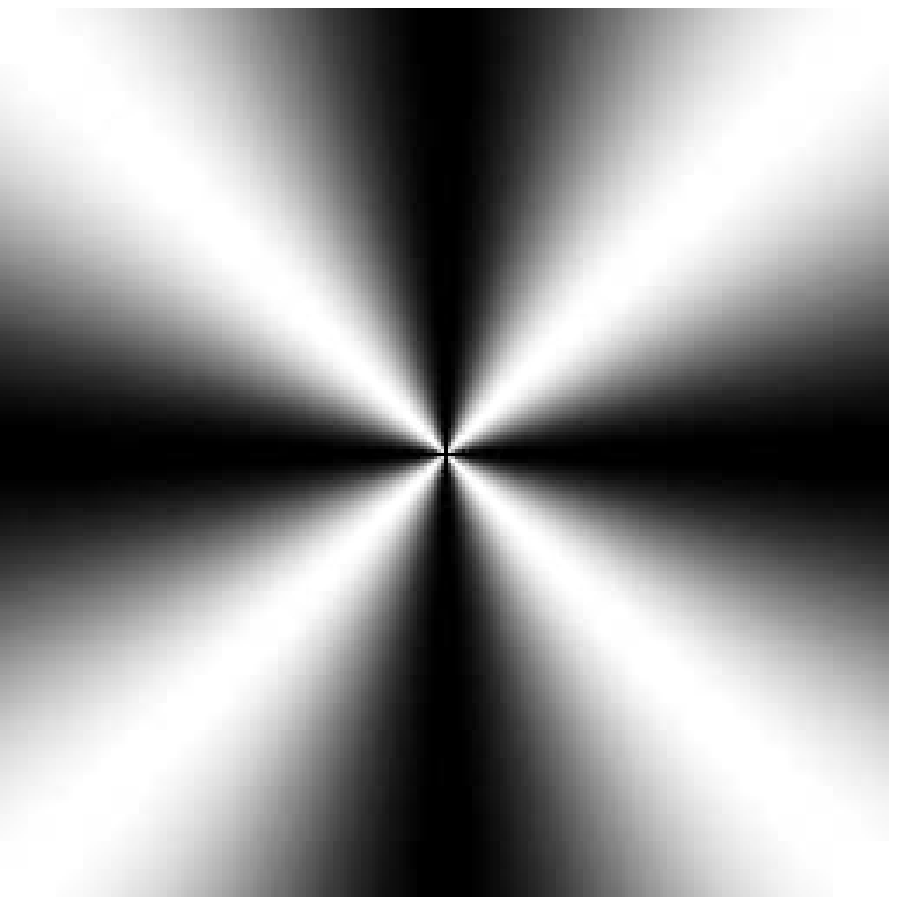}
  \caption{Particular example of an OVVC of topological charge $l_p=2$, illustrating the concepts needed to describe vectorial vortices. Left: the figure shows a HWP with an optical axis orientation $\omega$ that rotates about the center as $\omega=\theta$, where $\theta$ is the azimuthal coordinate (dashed lines perpendicular to the circumference). The textbook net effect of a HWP on a linear impinging polarization is to rotate it by $-2\times \alpha$ where $\alpha$ is the angle between the incoming polarization direction and the extraordinary ($e$) optical axis . Applying this property to the space-variant HWP, assuming an incoming horizontal linear polarization, we see that the latter is transformed by the vectorial vortex so that it spins around its center twice as fast as the azimuthal coordinate $\theta$ (arrows). In this case, the angle of local rotation of the polarization vector $\phi_p$ corresponds to the ``geometrical'' or Pancharatnam phase: upon a complete rotation about the center, $\phi_p$ has undergone a total $2\times 2\pi$ virtual phase ramp, which corresponds to the definition of an optical vortex of topological charge 2. Right: computation of the periodic modulation of the intensity transmitted by an OVVC between crossed polarizers.\label{ovvc2pol}}
\end{figure}

A vectorial vortex optical element possesses no structural phase ramp per se. It is simply a halfwave plate (HWP) in which the optical axes rotate about the center (see Fig. \ref{ovvc2pol}). In the following we will show that the rotationally symmetric spatial variation of the optical axes manipulates the transverse polarization state in such a way that a \emph{virtual} phase ramp is embedded to the so- called ``geometrical'' phase \cite[see Appendix]{Berry87}.

\subsection{Spatial variation of the optical axis orientation}

When the local birefringence characteristics of an optical component vary from point to point, it is said to be space variant. Space-variant birefringent optical elements (SVBOE) are extensively used as polarization control devices \cite{Gorodetski2005,Biener2006,Hasman2007}. The range of applications is manifold: polarimetry, laser-beam shaping, laser machining, tight focusing, particle acceleration, atom trapping, contrast enhancement microscopy, image encryption, etc. 

SVBOE can be represented by a function describing the spatial variation of the optical axis orientation (here extraodinary, $e$)
\begin{equation}
\textbf{e}(r,\theta)=\left[\cos{\left[\omega(r,\theta)\right]}\, \textbf{X}+\sin{\left[\omega(r,\theta)\right]}\, \textbf{Y} \right]
\end{equation}
where $r$, $\theta$ are the polar coordinates, while $X$ and $Y$ are the cartesian unit vectors. $\omega$ is the local direction of the optical axis frame with respect to $X$ (Fig. \ref{spvarsch}).
Let us now consider the case of the helical geometrical phase SVBOE. For a helical phase, the optical axis frame orientation is given by $\omega(r,\theta)=(l_p/2)\theta$, where $l_p$ is the so-called ``Pancharatnam topological charge'' (see Appendix). The optical axis frame function therefore becomes
\begin{equation}
\textbf{e}(r,\theta)=
\left[\cos{\left[(l_p/2)\theta \right]}\, \textbf{X}+\sin{\left[(l_p/2)\theta \right]}\, \textbf{Y} \right]
\end{equation}
We will demonstrate below that this spatial variation of the optical axis orientation is responsible for a helical \emph{phase structure} resulting from the geometrical phase, as opposed to the usual scalar phase.

\subsection{Optical vectorial vortex coronagraph}

A \emph{vectorial vortex} is a SVBOE with rotational symmetry properties. In other words, it is a halfwave plate in which the optical axes rotate about the center. As such it can be represented by a space-variant Jones matrix \cite{Born99}. Let us assume that $J_{HWP}$ is the Jones matrix for the equivalent spatially uniform halfwave plate, i.e.
\begin{equation}\label{svbe}
J_{HWP}(r,\theta)=\left[
\begin{array}{cc}
\eta_{o}e^{-i\Delta\phi_{o-e}/2} &0 \\
0 &\eta_{e} e^{+i\Delta\phi_{o-e}/2}
\end{array}
\right]
\end{equation}
where $\Delta\phi_{o-e}$ (for the sake of simplicity, $\Delta\phi$ in the following) is the $ordinary-extraordinary$ phase shift (also called the linear retardance), and where $\eta_{o}$ and $\eta_{e}$ are the local SVBOE transmittances along the $ordinary$ and $extraordinary$ directions of polarization, respectively.\footnote{Assuming no diattenuation, these transmittances can still be different because of the different Fresnel reflection coefficients resulting from the existence of the two different refractive indices that give birth to the birefringence.} For a perfect HWP, we have $\Delta\phi_{o-e}=\pi$, and $\eta_{o}=\eta_{e}=1$.

The resultant Jones matrix for a helical retarder is
\begin{equation}\label{vortex}
J_{v}(r,\theta)=R\left[\omega(r,\theta)\right]J_{HWP}(r,\theta)
R\left[\omega(r,\theta)\right]^{-1}
\end{equation}
where $R\left[\omega(r,\theta)\right]$ is the rotation matrix (which depends on $(r,\theta)$)

\begin{equation}\label{rot}
R\left[\omega(r,\theta)\right]=\left[
\begin{array}{cc}
\cos{\omega(r,\theta)} &-\sin{\omega(r,\theta)} \\
\sin{\omega(r,\theta)} &\cos{\omega(r,\theta)}
\end{array}
\right]
\end{equation}

For phase errors only ($\eta_{o}=\eta_{e}=1$), the final Jones Matrix is then 
\begin{equation}
J_{v}(r,\theta)  =  \sin{\frac{\Delta\phi}{2}}  \left[\begin{array}{cc}
\cos{l_p\theta} & \sin{l_p\theta}\\
\sin{l_p\theta} & -\cos{l_p\theta}\end{array}
\right]\\
   -  i \cos{\frac{\Delta\phi}{2}} \left[\begin{array}{cc}
1 & 0\\
0 & 1
\end{array}\right]
\end{equation}
or, assuming that $\Delta\phi=\pi+\epsilon$ with $\epsilon \ll 1$,
\begin{equation}
J_{v}(r,\theta)  = \left[\begin{array}{cc}
\cos{l_p\theta} & \sin{l_p\theta}\\
\sin{l_p\theta} & -\cos{l_p\theta}\end{array}
\right]\\
   +  i\frac{\epsilon}{2}  \left[\begin{array}{cc}
1 & 0\\
0 & 1
\end{array}\right]
\end{equation}
The rotating output field is represented by the first term (see Fig. \ref{ovvc2pol}) while phase deviations show up in the second matrix (and are polarized in parallel to the input field).

Although the analysis has so far been presented in the linear basis (see also \cite{Mawet2005a}), the beauty of the vectorial vortex theory reveals itself in the so-called helical basis, i.e.\ with right and left-handed
circular polarization unit vectors
 \begin{equation}\label{roteq1circR}
RC_{in}=\left[
\begin{array}{c}
1 \\
0
\end{array}
\right],\ LC_{in}=\left[
\begin{array}{c}
0 \\
1
\end{array}
\right]
\end{equation}
In the helical basis, the vortex Jones matrix must be transformed following
\begin{equation}\label{vortexcirc}
J_{v}(r,\theta)=U
J_{v}(r,\theta) U^{-1}
\end{equation}
With the so-called helical-basis transformation matrix\cite{Bomzon2002}
\begin{equation}\label{u}
U=\frac{1}{\sqrt{2}}\left[
\begin{array}{cc}
1 & i\\
1 & -i
\end{array}
\right]
\end{equation}
Substituting Eqs. (\ref{svbe}) and (\ref{rot}) into Eq. (\ref{vortex}), and then Eq. (\ref{vortex}) into Eq. (\ref{vortexcirc}) with Eq. (\ref{u}), finally yields the vectorial vortex Jones Matrix

\begin{equation}
J_{v}(r,\theta)  =  V \left[\begin{array}{cc}
0 & e^{il_p\theta}\\
e^{-il_p\theta} & 0
\end{array}\right]\\
   +  L\left[\begin{array}{cc}
1 & 0\\
0 & 1
\end{array}
\right]
\end{equation}
where $V = \frac{1}{2}(\eta_{o}e^{-i2\Delta\phi/2}-\eta_{e}e^{+i2\Delta\phi/2})$ and $L = \frac{1}{2}(\eta_{o}e^{-i2\Delta\phi/2}+\eta_{e}e^{+i2\Delta\phi/2})$ are weighing coefficients depending on the local imperfections of the HWP, i.e. the retardance, and polarization component transmittances. In the perfect case where the transmittances are equal (and equal to 1, which means 100$\%$ throughput), i.e. $\eta_{s}=\eta_{p}=1$, and the retardance equals to $\pi$, $\Delta\phi=\pi$, we obtain as respective outputs to the input vectors of Eq. (\ref{roteq1circR})
\begin{equation}\label{roteq1circb}
RC_{out}=\left[
\begin{array}{c}
0 \\
e^{i(l_p\theta)}
\end{array}
\right],\ LC_{out}=\left[
\begin{array}{c}
e^{-i(l_p\theta)} \\
0
\end{array}
\right]
\end{equation}
This result clearly demonstrates that the two output polarization beams bear the canonical vortex phase structure, while being orthogonal and therefore decoupled. Note that $J_{v}$ is composed of two main terms. The first term is the pure vortex term of weight $V$, bearing the geometrical phase ramp structure $e^{il_p\theta}$. This phase modification is geometrical, i.e. induced only by the space variation of the optical axis orientation across the component. Subsequently, it is \emph{achromatic} by nature. The second term with a weighing coefficient $L$ bears no phase modification and is a leakage term.  A very important property of vectorial vortices versus scalar vortices, is that the polarization structure of the chromatic leakage due to the local imperfections of the HWP is orthogonal to the pure achromatic vortex term: note the different directions of the two Jones matrices of $J_v$ for circular polarization. This important fundamental difference between the two approaches suggests that one might be able to reduce the effect of the manufacturing defects (retardance and transmittance mismatches) by appropriate polarization filtering. An input left (resp. right) circular polarizer yields a right (resp. left) - circularly polarized vortex and a left (resp. right) - circularly polarized leakage. An output right (resp. left) - circular analyzer would let the pure vortex out only, blocking the left (resp. right) - circularly polarized leakage.

We just rigorously showed that vectorial vortices, like scalar vortices, present the same screw phase dislocation $e^{i(l_p\theta)}$ that induces the central phase singularity. As natural unpolarized light can always be decomposed into two orthogonal and mutually incoherent polarization states, the vectorial vortex, using polarization manipulation, actually works in natural unpolarized light.

\subsection{Sensitivity to aberrations}

\subsubsection{Low-order aberrations}
The optical vortex coronagraph attenuation sensitivity to low-order aberrations like tip-tilt $t$ (pointing error) has been shown, numerically by \cite{Mawet2005a} and \cite{Guyon2006}, and analytically by \cite{Jenkins2008} to be proportional to $t^{l}$. Leakage due to the finite size of stars can then be very easily estimated using a model of a uniform stellar disk which is a superposition of incoherent point sources. Assuming for instance an angular size of $0.01 \lambda/d$ ($\lambda$ is the observation central wavelength and $d$, the telescope diameter), one get a total leakage (integrated over the star) of $8\times 10^{-5}$ for a topological charge $l=2$, $10^{-8}$ for $l=4$, $10^{-12}$ for $l=6$, respectively.

\subsubsection{Chromaticity}\label{chrom}
As far as the sensitivity to the bandwidth (or chromaticity) is concerned, as every other phase mask coronagraphs such as the FQPM, it is one of the most difficult property to tackle. It is very dependent on the technological approach chosen to render the phase helix (see Sect. \ref{manuf}). For the vectorial vortex, chromatic effects appear in the wavelength dependence of the local polarization characteristics of the birefringent medium, i.e. the retardance and the ordinary and extraordinary transmittances. The consequence is that the weighing coefficient $V$ and $L$ are wavelength dependent: $V(\lambda)$ and $L(\lambda)$. To improve the achromaticity of the vectorial vortex, the first obvious solution is to make the polarization characteristics immune to wavelength changes. Multi-layer LCP designs, using dispersion compensation between layers, a technique equivalent to the making of achromatic waveplates \cite{Mawet2006}, are currently investigated as a potential solution. A one-layer design (our prototype) yields a $\sim$0.1 radian rms retardance error over a 20\% bandpass. A two-layer design yields a $\sim$0.01 radian rms over the same bandpass. Three layers would in theory allow us to reach the $\sim$0.001 radian rms level, etc. Neglecting the additional technical difficulties induced by the stacking, each additional layer therefore yields a 100-fold improvement in terms of performances (see Sect. \ref{limitations} and Eq. (\ref{nullingform})). Another solution is to filter the leakage term, as already mentioned. This solution, which could potentially bring $\sim$1000-fold improvement with on-the-shelf polarization components, has already been investigated theoretically elsewhere \cite{Mawet2007}. Experimental validation of this filtering method will be the subject of future work.

In any case, phase control is always more difficult than amplitude control. Making achromatic or pseudo-achromatic vortices is at the center of the current R\&D efforts in the field, and will be discussed further in the following (Sect. \ref{discussion} and Sect. \ref{perspectives}).

\section{Manufacturing an OVVC2 prototype using Liquid Crystal Polymers}\label{manuf}

\begin{figure}[!t]
  \centering
\includegraphics[scale=0.6]{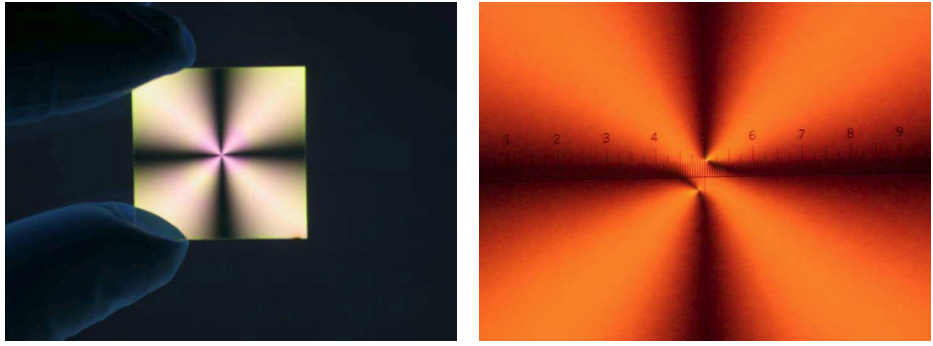}
  \caption{Pictures of the $l_p=2$ LCP device. Left: global picture taken between crossed polarizers (size = 2 inch squares). Right: zoom under polarimetric microscope at 100x magnification, showing a superimposed graduated scale (1 unit = 100 microns).\label{samp}}
\end{figure}

Several techniques exist to implement an OVVC. One can think of using natural birefringent crystals, but this is obviously not convenient to generating the desired smooth birefringent space-variant distributions\footnote{Let us mention the example of the four-quadrant phase-mask coronagraph made out of halfwave plates as a discrete optical vectorial vortex of topological charge 2 \cite{Mawet2006}. This technology has been chosen to make the FQPMs of SPHERE, the extreme adaptive optics second-generation instrument dedicated to exoplanet finding at ESO's Very Large Telescope \cite{Beuzit2007,Boccaletti2008}}. The second possibility is to generate artificial birefringence using a subwavelength grating so-called form birefringence. This technique has extensively been described elsewhere \cite{Mawet2005a, Mawet2005b}. It is to be noted that this technology is mostly adapted to longer wavelengths where the fabrication constraints are relaxed since the subwavelength grating feature line is proportional to the wavelength \cite{Niv06a,Hasman2007}. Several prototyping operations are underway using this technique. But once again, at short wavelengths, this technique is very difficult, because of the large aspect ratio of the grooves. Here we focus on a new approach involving a very promising technology based on liquid crystal polymers \cite{McEldowney2008a,McEldowney2008b}.

\subsection{LCP technology and fabrication technique}\label{fab}

Liquid crystal polymers are materials that combine the birefringent properties of liquid crystals with the excellent mechanical properties of polymers. The orientation of a LCP is achieved through photo-alignement. Once aligned and cured, the polymer reaches a very stable solid state. A description of the specific proprietary process used by JDSU Uniphase (formerly Optical Coating Lab, Inc.) to make the LCP vortices follows. A photo-alignment layer (ROP108 from Rolic) is spin-coated onto the substrate, baked, and then the alignment is set through exposure to linear polarized UV (LPUV) light. The alignment layer is exposed through a narrow wedge shaped aperture located between the substrate and the polarizer. Both the polarizer and the substrate are continuously rotated during the exposure process in order to create a continuous variation in photo-alignment orientation as a function of azimuthal location on the substrate. The variation in fast axis orientation with azimuthal angle is
determined by the relative rotation speeds. The LCP precursor (ROF5104 from Rolic) is then spin-coated and subsequently polymerized using a UV curing process. It naturally orients itself according to the photo-alignment layer. A post-bake stabilizes the films into a solid SVBOE.

Using this approach, we fabricated a prototype LCP device on 2 inch squares of fused silica glass (Fig. \ref{samp}). A broadband anti-reflective coating was applied on the back surface. The LCP single-layer coating was then laminated with an optical adhesive and glued to a second fused silica substrate having an anti-reflective coating on the opposite side. We targeted a half wave retardance for wavelengths near 1650 nm, in the center of the astronomical H band. For the sake of simplicity, our first sample was chosen to be a single-layer design. For this reason, it is relatively chromatic by construction (see Sect. \ref{results}).

The prototype was produced with a topological charge 2. This particular topological charge was chosen for several reasons: first, it provides a very small inner working distance coronagraph (useful to reduce aperture sizes), second, it is technologically easier to produce as a first try. Other samples that meet more stringent specifications and higher topological charges are under production. Fig. \ref{samp} shows our sample between crossed polarizers. The resultant transmission pattern conforms well to the theoretical analysis in linear polarization presented in Fig. \ref{ovvc2pol}.

\subsection{Sample polarimetric characterization}\label{carac}

\begin{figure}[!t]
  \centering
  \includegraphics[scale=0.3]{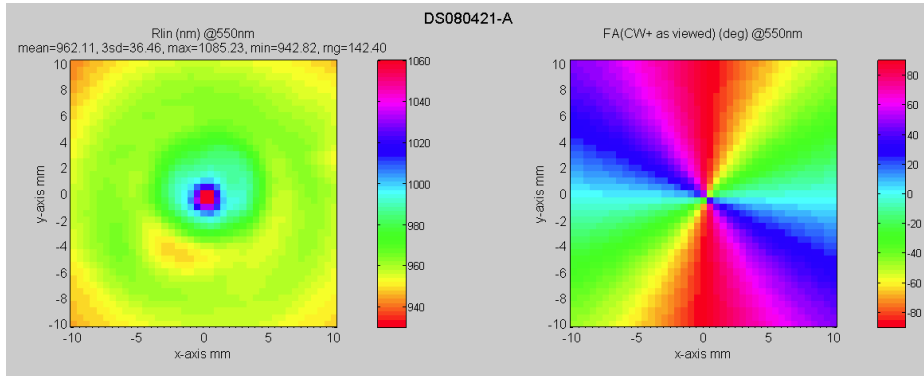}
  \caption{Axoscan sample characterization at $\lambda=550$ nm. Left: linear retardance in nm. Right: fast axis orientation. Taking the LCP retardance dispersion behavior into account, the measured retardance value of 962 nm at $\lambda=$550 nm is found to be equivalent to the target retardance of 826.5 nm at $\lambda=$1650 nm (halfwave). The retardance map suggests a small relative rise in retardance at the center region.\label{carsamp}}
\end{figure}

The polarization properties of our first prototype were mapped on an Axoscan Mueller Matrix Spectro-Polarimeter available from Axometrics. The sample was placed on an x-y stage and mapped at a 0.5x0.5 mm xy-resolution over a 20x20 mm area. The beam size was about 2 mm, so the measurement at each point represents an average over the beam area. For our $l_p=2$ sample, summary maps of (a) linear retardance; (b) fast axis orientation are provided in Fig. \ref{carsamp}. The fast axis maps are given in degrees and with a range from -90 to 90 degrees. The orientation in all cases varies continuously. Those results showed excellent match to the specifications.
 \begin{figure}[!b]
  \centering
 \includegraphics[scale=0.17]{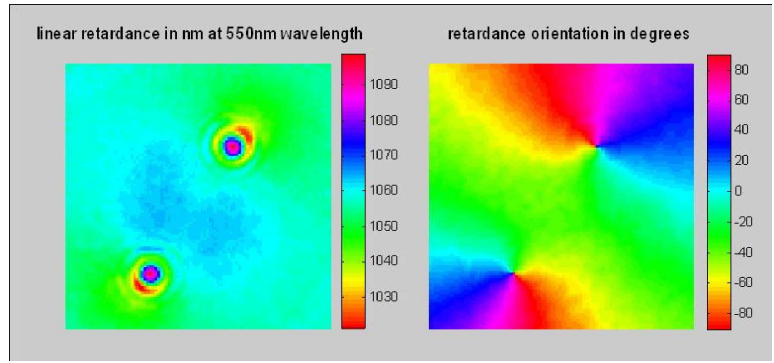}
  \caption{Magnified image of the linear retardance in nm (left) and retardance orientation (right) for the first prototype measured using a Mueller Matrix microscope. The field of view is 100 x 100 microns. \label{highres}}
\end{figure}

The first prototype samples were then measured on a Mueller Matrix microscope to determine the specific polarization properties of the sample close to the center of the vortex phase mask at ultra high resolution. The system used to make these measurements consists of a Mueller Matrix imaging polarimeter setup such that the microscope and camera field of view was approximately 100 microns \cite{Pezzaniti95}. The instrument is located at the University of Arizona College of Optical Sciences polarization laboratory. The CCD detector active area was 100x100 pixels so the instantaneous field of view of each pixel was 1 micron. The measurement wavelength was 550 nm matching that of the Axoscan polarimeter.

The measurement of the retardance and retardance orientation is given in Fig. \ref{highres}. The retardance measurement at 550 nm indicates the retardance is uniform except at two points symmetrically spaced about the center at radii of $\sim 34$ $\mu$m, where the orientation spatial variation is large. The value of the retardance is very close to that measured by the Axoscan at the center of the vortex retarder. The retardance orientation measurement indicates that the orientation is continuous at the 1$\mu$m resolution. The measurements also show that the orientation is discontinuous at the same 2 points close to the center. The region between these two points defines a central zone which deviates from the overall large scale pattern (the ``region of disorientation''). The distance between these points is approximately 68 $\mu$m. These two points are believed to originate from the manufacturing process. As described earlier, the process involves the rotation of the substrate and a polarizer (Sect. \ref{fab}). Both are rotating interlaced with the aperture and the UV lamp. The two singularity points defining the central region of disorientation are thought to be the footprints of a tiny misalignment of the rotation axes of those two moving mechanical parts. Tackling this misalignment will be our number one priority in the future.

\section{Coronagraphic Lab measurements}
We next tested the use of this LCP OVVC2 device as a coronagraphic mask. After describing the optical setup, acquisition and data processing, we discuss the measurement results in the light of the previously characterized limitations of this first prototype.

\subsection{Optical Setup}
\begin{figure}[!t]
  \centering
 \includegraphics[scale=0.6]{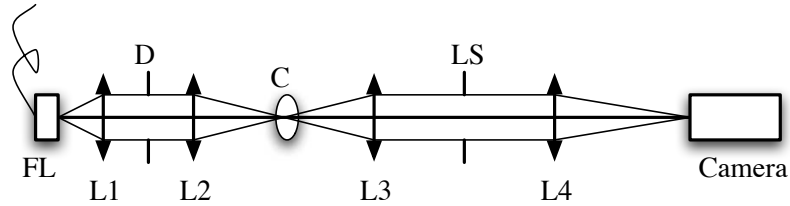}
  \caption{Coronagraphic transmission bench. Light from the fiber-coupled sources is injected via the fiber launcher (FL), and is collimated by the lens L1. We define the pupil with a iris diaphragm D, before focussing with L2 onto the coronagraphic mask (C). L3 reimages the entrance pupil where diffraction from the coronagraphic mask is filtered out thanks to the Lyot stop (LS). L4 makes the final image on our Merlin InSb camera.\label{bench}}
\end{figure}

The optical setup (Fig. \ref{bench}) is a near-infrared coronagraphic transmission bench already used at JPL for previous tests of four-quadrant phase-mask coronagraphs \cite{Haguenauer2005}. To simulate starlight, we used fiber-coupled sources: either a laser diode at 1550 nm, or a white light provided by a halogen lamp at a temperature of 3400 K (10 Watts). We used standard telecom monomode fibers to provide a clean spatially filtered wavefront as input. We used standard AR-coated achromatic near-IR lenses with low frequency errors ($<\lambda/10$ PTV over $10\, mm$ in diameter). The entrance pupil is defined by an iris diaphragm that allows us to easily vary the $F$ number of the beam ($F=f/d$, with $f$ the focal length of the imaging lens, and $d$ the diameter of the beam as defined by the entrance iris diaphragm). To filter the post coronagraphic mask pupil (Lyot) plane, we use an iris diaphragm with a diameter set to be smaller than the entrance pupil by about $20\%$. 

\subsection{Image acquisition and data processing}
All images were recorded with a LN2 cooled 12-bit Merlin InSb camera. The cold filter in the camera is a Spectrogon shortwave pass filter (cut-off around 2.6 microns). We also used a JDSU H-band room temperature filter to define a 12\% bandpass around 1650 nm when using the white light. We oversampled the images relative to $F \lambda$ to allow a proper peak-to-peak attenuation measurement. To assess the coronagraphic performance of the component, we first built up a coronagraphic image by adding 100 individual exposure of 16 ms. The number of frames and the individual exposure time are chosen to balance signal with background and read-out noise. The next step was to record direct non-coronagraphic images (the mask was translated off the focal plane spot) in the same conditions. All images had a median dark frame subtracted from them. The instantaneous single exposure dynamic range of this particular 12-bit camera is typically limited to a few $10^{-4}$.

\section{Results}\label{results}

\subsection{Metrics}
In practice, one can quantify a coronagraph's ability to attenuate light with different figures of merit, for example:
\begin{itemize}
\item[-] the total rejection $R$ (or its inverse, the total attenuation $N$): ratio of total intensity of the direct image to that of the coronagraphic image;
\item[-] the peak to peak (PTP) rejection (or its inverse, the PTP attenuation): ratio of the maximum intensity of the direct image to the maximum intensity of the coronagraphic image;
\item[-] or more generally, the contrast as a function of radial offset from the center (in $\lambda/d$ units).
\end{itemize}

It is important to note that these metrics are not always related in the same way according to the working conditions. The first one is representative of the averaged attenuation of all the starlight impinging in the coronagraphic system. This value is often difficult to measure with great accuracy, since noise interferes with its evaluation far from the central peak, where the signal is feeble. For that reason, we also use here the PTP rejection which is easier to measure with a high SNR. Peak-to-peak rejection results are often better than the expected total rejection because of the transformation of the PSF structure by the coronagraph. The PTP rejection has its practical importance though, because in an actual observing instrument, it dictates the maximum integration time allowed by the central peak saturation.

\subsection{A priori known limitations of the first prototype} \label{limitations}

The sample was designed to provide a $\pi$  radian phase shift at the central wavelength of the astronomical H band, $\sim 1.65$ $\mu$m. The natural retardance dispersion of our single layer LCP sample over the $12\%$ bandpass defined around the central 1.65 $\mu$m wavelength is $\sigma \approx 0.115$ radian rms, where $\sigma$ is the retardance standard deviation over the filter bandpass. According to the phase mask nulling formula \cite{Mawet2006}, 
\begin{equation}\label{nullingform}
N=R^{-1}=\frac{\sigma^2}{4} 
\end{equation}
this value yields a nominal broadband rejection ratio of $R\approx 300$. On the other hand, the laser diode we used operates at 1550 nm, which induces for the same reason a retardance error $\epsilon \approx 0.21$ radian, limiting the total rejection ratio to $\sim 90$. Indeed, the nulling formula is identical in the polychromatic and monochromatic case, expect for the substitution of the rms error $\sigma$ with the  error $\epsilon$.
\begin{table}[!b]
\begin{center}\caption{Expected and measured total rejection factor (R) as a function of the defect diameter size (relative to $F\lambda$). The measured peak-to-peak rejection ratio (PTP) is also given. The first set of measurements were made with the 1550 nm laser diode, and the last three lines  with broadband (BB) light, i.e. taken with the halogen white light source and a 12\%-bandwidth H-band filter.\label{table1}}
\begin{tabular}{c|cccc}
\hline\hline
 &Centr. def. diam./$F\lambda$ & Exp. R &Meas. R & Meas. PTP \\
 \hline
&$1.12$ &2 &3.8 &12.6 \\
&$0.56$ &4 &8 &21 \\
Laser &$0.28$ &12 &20 &50\\
&$0.14$ &34 &36 &57\\
&$0.10$ &46 &42 &70\\
\hline
&$1.12$  & 2 &3.7 & 9\\
BB & $0.56$  & 4 & 7.5 & 16.2\\
&$0.28$  & 13.5 & NA &30 \\
\hline
\end{tabular}
\end{center}
\end{table}

Unfortunately, the microscope image (Fig. \ref{samp}, right) shows a central region of disorientation of about 68 $\mu$m in diameter, and the central region is the most important one in coronagraphic applications since most of the stellar energy to be suppressed will fall on it. Despite the complexity of the central region of disorientation (two vortex centers are visible), we can consider it as a mere hole. This overly conservative hypothesis yields a very strong limitation on the rejection as synthesized in Table \ref{table1}, second column. Indeed, most of the stellar peak energy falls within the beam FWHM, $F \lambda$. The total rejection, R, should be approximately

\begin{equation}\label{model}
R \approx \left[e\left(\frac{s}{F\lambda}\right)+\frac{\sigma^2}{4}\right]^{-1}
\end{equation}

where $e(\frac{s}{F\lambda})$ is the encircled energy function at the angular separation $\frac{s}{F\lambda}$, $s$ being the defect size (central region of disorientation). $\sigma$ (radian rms) is the standard deviation of the retardance owing to the LCP dispersion over the working wavelength range. It can be substituted with $\epsilon$, the retardance error (radian) at the single working wavelength.

\subsection{Pupil imaging and vortex creation} 

\begin{figure}[!t]
  \centering
  \includegraphics[scale=0.5]{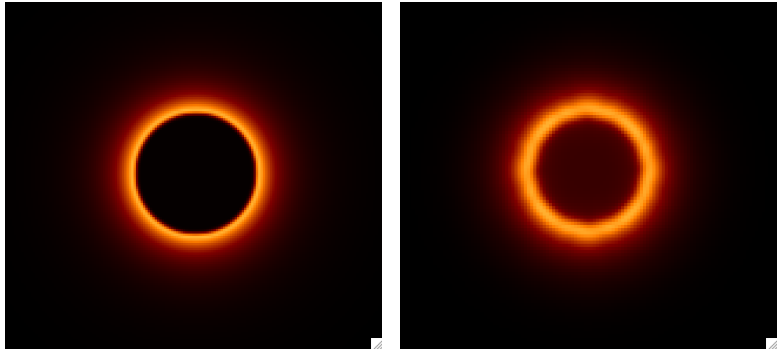}
  \caption{Left: theoretical pupil image after the vortex showing the rejection of the input on-axis light inside the geometric pupil area in a circularly symmetric shape. Right: experimental image. The stretch is linear in both images.\label{pupil}}
\end{figure}

Our coronagraphic bench is also equipped with a pupil imaging mode.This mode allows us to reimage the Lyot pupil plane (entrance pupil seen through the coronagraph). Using this mode reveals the vortex effect, predicted by the theory (Fig. \ref{pupil}, left) and providing the coronagraphic property of the vortex (Fig. \ref{pupil}, right). 

\subsection{Coronagraphic profile measurements} 

\begin{figure}[!t]
  \centering
\includegraphics[scale=0.3]{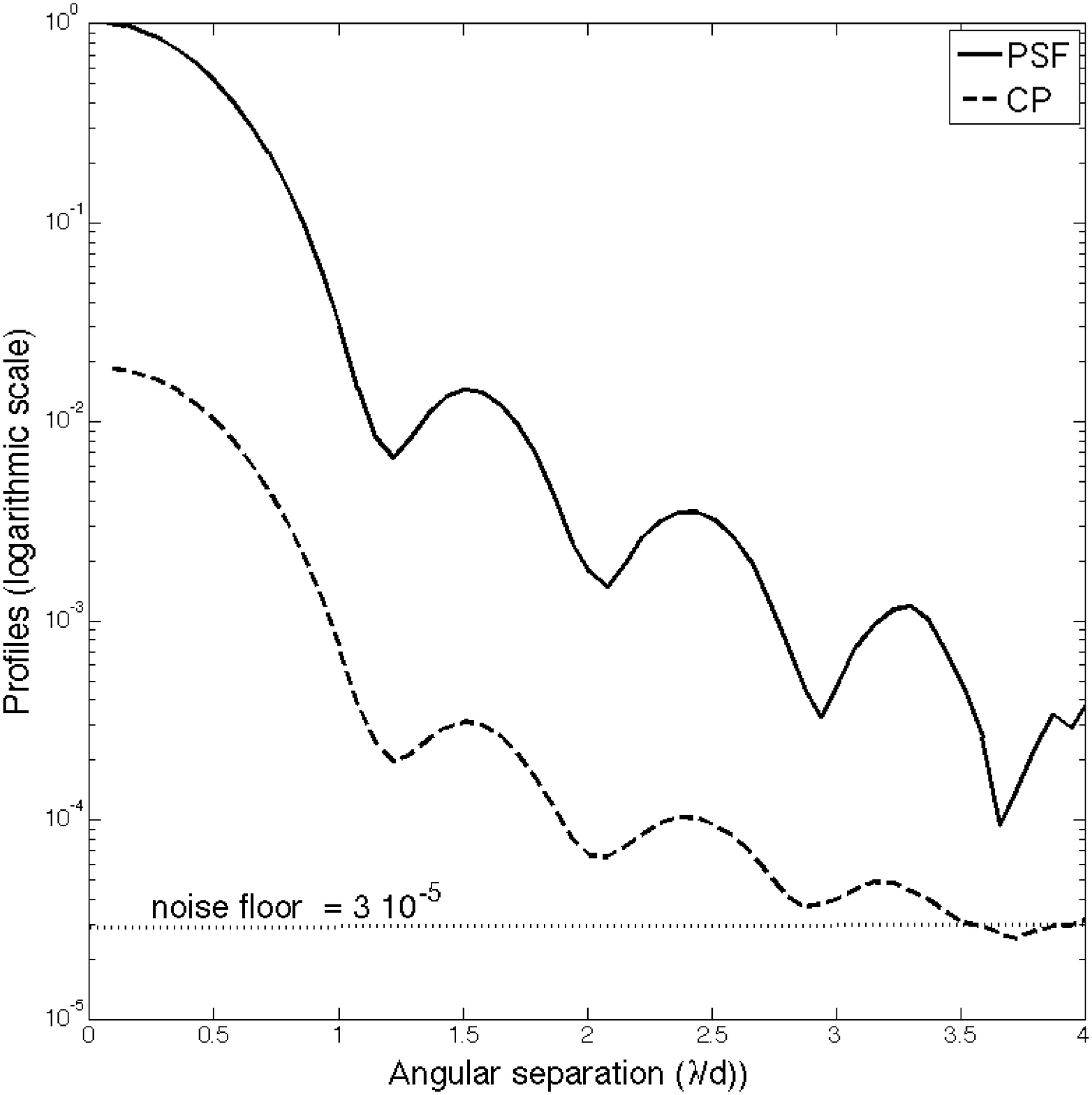}
  \caption{Azimuthally averaged coronagraphic profile (CP) obtained with the OVVC2 using LCP, measured at 1550 nm and with a virtual defect size $10\%$ of the beam size $F\lambda$. The PSF is also shown in dashed line. The camera noise floor (dotted straight line) of about $3\times 10^{-5}$ is quickly reached at 3.5 $\lambda/d$\label{cp}.}
\end{figure}

Being aware of the potential harmful effect of the central region of disorientation, we measured focal plane point-spread functions at different $F$'s. This allowed us to mitigate this first try technological defect, by changing the defect size relative to the beam FWHM $F\lambda$. Results are summarized in Table \ref{table1}. The first set of measurements was done with the 1550 nm diode laser. The last three lines of the table show results obtained with the white light source and the 12\% H-band filter (centered at 1650 nm). In Fig. \ref{cp}, we also show the coronagraphic profile obtained with the 1550 nm diode in the most favorable condition, showing a peak-to-peak rejection of  $\sim 70$. A contrast of $3\times 10^{-5}$ has been obtained at the camera noise floor in the corresponding working conditions, which is attained at $\sim 3.5$ $\lambda/d$.

\section{Discussion}\label{discussion}

\begin{figure}[!t]
  \centering
  \includegraphics[scale=0.3]{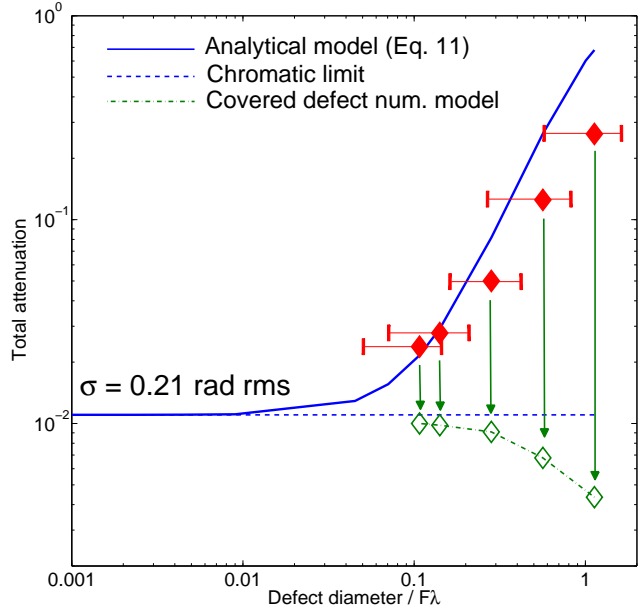}
  \caption{Log-log graph of the total attenuation versus relative central defect size diameter. The dashed line gives the limit set by the chromaticity (or phase error at a single wavelength), the solid one the expected total attenuation given by the analytical model of Eq. (\ref{model}), the dash-dotted one is linking data points (hollow diamonds) resulting from the numerical modeling of the effect of the defect coverage by an opaque mask whose size is matched to the region of disorientation. Our uncertainty for the diameter of the central region of disorientation to use (owing to our simplistic hole model) is represented by the horizontal error bars on the measurement points.\label{perf1}}
\end{figure}

The results from Table 1 are synthesized in Fig. \ref{perf1}, where the experimental data (filled diamonds) are plotted along with the analytical model of Eq. (\ref{model}) (solid lines), and the chromaticity limit given by Eq. (\ref{nullingform}) according to the single-layer design (dashed line). The most important conclusion evident in Fig. \ref{perf1} is that our simple analytical model results for the performance of the coronagraph (Eq. (\ref{model})), based on the identified central defect of the prototype and chromaticity, approximately matches the measurements (Table \ref{table1}) to the first order. 

The best results at 1550 nm (Fig. \ref{cp}) are near the limit set by the residual chromatism discussed above. Results in white light are consistent with the model of Eq. (\ref{model}), even if limited by photon noise because of the low power of a single mode white light source. The bench will soon be upgraded with a supercontinuum white light source to enable higher contrast observations.

Next, we consider what happens if we cover the central region of disorientation of our prototype with an opaque mask. To precisely assess the diffraction effect that this would induce, numerical simulations with PROPER were used. PROPER is an optical propagation library for IDL developed by John Krist \cite[available from www.openchannelsoftware.com]{Krist2007}. It allows us to model the propagation through a realistic model of the coronagraphic mask using Fourier transforms, from the entrance pupil to the final image plane. It is to be noted that those numerical simulations also confirm the Eq. (\ref{model}) model, in the uncovered and chromaticity-limited regimes.

One very interesting outcome of these simulations of the covering or \emph{hybridization} of the phase mask with a very simple amplitude profile, is that it can help us to mitigate the chromaticity (see the dash-dotted curve linking hollow diamonds in Fig. \ref{perf1}). Covering the central region of disorientation thus not only helps us to reach the chromatic limit, but also to beat it significantly, as represented by the vertical arrows in Fig. \ref{perf1}. Therefore, by covering the central defect, this vortex mask could be made which meet the requirements of current ground-based coronagraphy (total attenuation of a few hundreds), set by residual wavefront errors after adaptive optics correction of atmospheric turbulence.

\section{Perspectives}\label{perspectives}

Our simple model (Eq. (\ref{model})), which has been confirmed to first order by measurements (Fig. \ref{perf1}), can be used to predict future performance as a function of error terms in Eq. (\ref{model}), under the assumption that potential second-order defects which are currently not identified can be ignored. Note that for small defect, the rejection is
\begin{equation}\label{model2}
R \approx \left[\left(\frac{s}{F\lambda}\right)^2+\frac{\sigma^2}{4}\right]^{-1}
\end{equation}
If we set a rejection goal, we can conversely derive the following individual constraints, since each term must be smaller than $\frac{1}{R}$, we have
\begin{equation}\label{model3}
\frac{s}{F\lambda} < \frac{1}{\sqrt{R}} 
\end{equation}
\begin{equation}\label{model4}
\sigma <\frac{2}{\sqrt{R}}
\end{equation}

\begin{figure}[!t]
  \centering
 \includegraphics[scale=0.3]{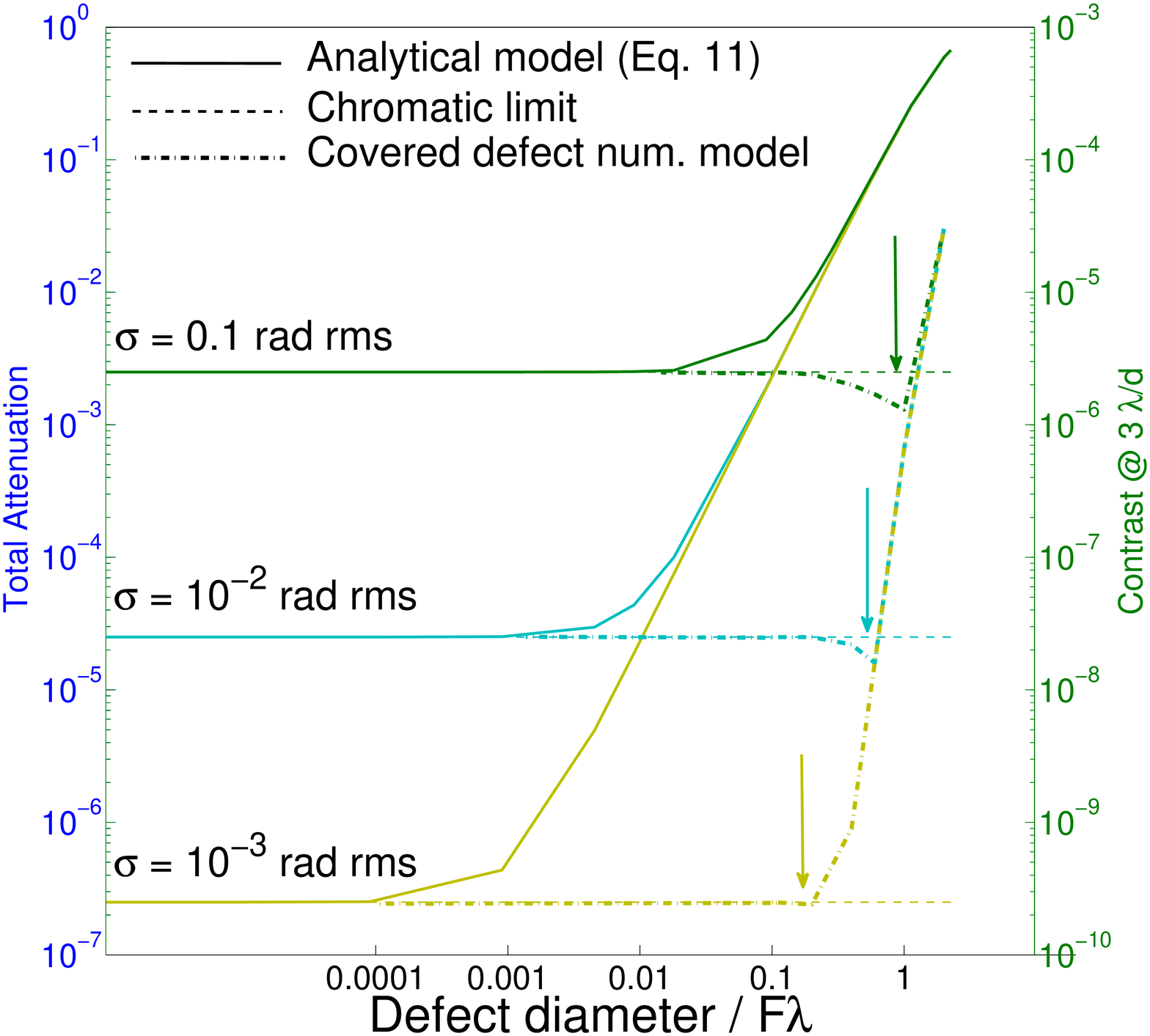}
  \caption{Performance prediction based on analytical and numerical modeling of the vortex, extrapolated to very high contrast levels, ignoring unknown second-order effects. The vertical arrows shows the expected gain obtained by masking the central defect region with an opaque mask to reach the limit set by residual chromatism. \label{perf2}}
\end{figure}

However, in general, these two terms must be combined as in Eq. (\ref{model}), and as plotted in the solid curves of Fig. \ref{perf2}. Once again, we next considered covering up the central region of disorientation by an opaque mask. The main results of the optical propagation analysis with PROPER is that by covering up the central zone, the device can once again operate at the chromatic limit for much larger central defects (Fig. \ref{perf2}, dash-dotted line). Thus, by covering up the central defect, the constraint of Eq. (\ref{model3}) can be dramatically relaxed. Our numerical simulations using PROPER also show that the size of the occulting zone ought to be within $\sim 1\lambda/d$ in diameter, beyond which too much of the vortex is covered.

\section{Conclusions}

We have demonstrated for the first time that a vortex synthesized out of a space-variant birefringent element, i.e. a vectorial vortex acts as a viable coronagraph, as predicted. Thanks to ground-breaking advances in LCP technology, a usable sample was produced in a very short time. Even though the first prototype falls short of the specifications of next generation ground-based instruments for exoplanet imaging and characterization, such as SPHERE \cite{Beuzit2007}, GPI \cite{Macintosh2007}, Palm-3000 \cite{Dekany2006}, and HiCIO \cite{Tamura2006}, we have easily identified the major Achille's heel of the prototype, the central region of disorientation. We have good reasons to believe that this purely mechanical problem can be overcome (Sect. \ref{carac}). It is currently being addressed, with the objective of reducing this region from 68 microns to a few microns. At such scales, as seen above, a simple occulting mask should be sufficient to hide the remaining defects without affecting the vortex effect down to the levels required by Earth-like planet detection.

The ability of LCP to synthesize optical vortices is excellent, with the potential of high-level achromatization and the straightforward possibility of implementing very high topological charges (samples of topological charge of 8 have already been produced, Scott McEldowney, personal communication). Prototyping operations with the purpose of assessing this potential in detail are underway at JPL with our industrial partner JDSU, both in the visible wavelength range on the High Contrast Imaging Testbed (HCIT), in the context of a small space-based observatory dedicated to exoplanet imaging \cite{Trauger2007}, and in the frame-work of extreme adaptive optics preparatory experiments from the ground \cite{Serabyn2007}.

\section*{Acknowledgments}
This work was carried out at the Jet Propulsion Laboratory (JPL), California Institute of Technology (Caltech), under contract with the National Aeronautics and Space Administration (NASA). The first author was supported by an appointment to the NASA Postdoctoral Program at the JPL, Caltech, administered by Oak Ridge Associated Universities through a contract with NASA. 
\appendix
\label{appendix}

\section*{Appendix: Pancharatnam or Geometrical Phase, topological charge}
\setcounter{equation}{0}
\renewcommand{\theequation}{A{\arabic{equation}}}
The Pancharatnam phase, or geometrical phase was discovered by Pancharatnam in 1955 \cite{Berry87}. Pancharatnam showed that a cyclic change in the state of polarization of light is accompanied by a phase shift determined by the geometry of the cycle as represented on the Poincar\'e sphere \cite{Born99}. 

The Geometric phase can be defined \cite{Berry87} as the argument of the inner product of the two Jones vectors describing the light beam whose state of polarization is made to change 
\begin{equation}\label{pancha1}
\phi_p=arg\langle E(r,\theta), E(r,0) \rangle
\end{equation}

In scalar fields, topological charges are defined according to the number of $2\pi$ radians that the phase of the scalar wave accumulates along a closed path surrounding a phase singularity. The topological charge of the vectorial vortex is defined in the same manner, but uses the Pancharatnam phase instead of the scalar phase
\begin{equation}\label{pancha2}
l_p=\frac{1}{2\pi}\oint \nabla \phi_p ds
\end{equation}
where the integration path encircles the phase singularity.

\end{document}